# Light-matter interactions in layered materials and heterostructures: from moiré physics and magneto-optical effects to ultrafast dynamics and hybrid meta-photonics

Luca Sortino[1], Marcos H. D. Guimarães[2], Alejandro Molina-Sánchez[3], Jiamin Quan[4], Denis Garoli[5,6], and Nicolò Maccaferri[7]

[1]Chair in Hybrid Nanosystems, Faculty of Physics, Ludwig-Maximilians-Universität München, 80539 Munich, Germany
[2]Zernike Institute for Advanced Materials, University of Groningen, The Netherlands
[3]Institute of Materials Science (ICMUV), University of Valencia, Catedrático Beltrán 2, E-46980, Valencia, Spain
[4]Institute of Precision Optical Engineering, School of Physics Science and Engineering, Tongji University, Shanghai, China
[5]Istituto Italiano di Tecnologia, Via Morego 30, 16136 Genova, Italy
[6]Università degli studi di Modena e Reggio Emilia, Dipartimento di Scienze e Metodi dell'ingegneria, Via Amendola 2, Reggio Emilia, Italy
[7]Department of Physics, Umeå University, Linnaeus väg 24, 901 87 Umeå, Sweden

## Abstract
Layered two-dimensional (2D) materials have revolutionized how we approach light-matter interactions, offering unprecedented optical and electronic properties with the potential for vertical heterostructures and manipulation of spin-valley degrees of freedom. The discovery of moiré physics in twisted heterostructures has further unlocked new possibilities for controlling the band structure of tailored semiconductor heterostructures. In parallel, the integration of 2D materials with hybrid photonic structures and ultrafast studies on their optical and spin-valley properties has revealed a wealth of novel physical phenomena. This perspective highlights the recent advances in our understanding of light-matter interactions in moiré and 2D systems, with a particular emphasis on ultrafast processes and the integration of these materials into photonic platforms. We explore the implications for optoelectronics and emerging photonic technologies, positioning 2D materials as a transformative tool for next-generation devices.

# 1. Introduction

Two-dimensional (2D) materials have captivated scientific interest over the past two decades due to their unique physical properties and potential for groundbreaking applications across optoelectronics and quantum technologies [1,2]. Emerging as atomically thin layers, these materials exhibit behaviors distinct from their bulk counterparts, largely owing to their reduced dimensionality leading to strong quantum confinement and distinctive symmetry properties. Among the diverse class of 2D materials, semiconductors such as transition metal dichalcogenides (TMDs), black phosphorus, and hexagonal boron nitride have revealed a wealth of novel phenomena, pushing the boundaries of material science and nanotechnology. In particular, TMDs have garnered attention for their remarkable properties when compared to conventional semiconductors. The strong Coulomb interactions, resulting from reduced dielectric screening, lead to tightly bound excitons with significant (>200 meV) binding energies, enabling high-speed processes on femtosecond timescales and strong light matter interaction [3]. These stable excitons, present also in bulk crystals and extremely enhanced in their 2D form, dominate the optical and charge transport properties in TMDs with favorable applications in photonics and opto-electronics[4,5].

Moreover, a direct band-gap transition in single layer promotes a strong light matter interaction, as, for example, TMDs can absorb to 10% of light in <1 nm thick layers. Additionally, excitons in 2D TMDs are affected by the reduced dielectric screening of the dielectric environment, exhibiting also stable Rydberg excitons, and single photon emission at cryogenic temperatures. Many types of multi-particle excitons are found in TMDs, such as trions, i.e. charged excitons, and biexcitons. Moreover, the development of van der Waals heterostructures of multiple TMDs layers give rise to new excitonic species, such as indirect or quadrupolar excitons, which expand the library of solid-state matter excitation that can be investigated [6]. Overall, the excitonic physics in 2D semiconductors is characterized by rapid formation and relaxation timescales, often on the order of femtoseconds to picoseconds, enabling ultrafast excitonic processes critical for high-speed applications [7].

Another peculiar feature that sets 2D semiconductors further apart from traditional materials is the coupled spin-valley physics [8,9]. In monolayer TMDs, the lack of inversion symmetry and strong spin-orbit coupling lead to a coupling of spin and valley degrees of freedom, resulting in two distinct valleys (K and K') in the band structure that are energetically degenerate, but spin-polarized. This coupling provides a unique platform for valleytronic applications, where information can be encoded and manipulated using valley indices. Spin-valley locking enables optically induced valley polarization and long-lived valley coherence, creating exciting opportunities for data storage and quantum information processing. This control over the spin-valley properties can become even more interesting when exploring the ultrafast dynamics properties of layered materials [10–13].

An additional layer of complexity and interest arises when 2D materials are stacked vertically, forming so-called van der Waals heterostructures [14]. The lack of covalent bonds between 2D layers allows additional freedom in designing device architectures, playing with number of layers, thickness and composition. More recently, the field of condensed matter physics have been shaken by the realization that by introducing a slight twist angle between adjacent layers, the physics of the 2D heterostructure can be changed profoundly, owing to the formation of moiré superlattices [6,15–17]. These superlattices introduce a periodic potential, on a scale of 1-100 nm, that modifies the electronic band structure, giving rise to flat bands, correlated electronic states, and tunable optical properties. In TMD heterostructures, moiré patterns facilitate novel light-matter interaction mechanisms, including the formation of moiré excitons, bound states that are spatially localized within the nm-long periodic moiré lattice. Such configurations enable exploration of excitonic Hubbard models, where interlayer excitons interact and form complex quantum phases, opening pathways to simulate strongly correlated electronic systems and study phenomena such as excitonic insulators and Mott physics [18]. Additionally, the tunability of moiré patterns by interlayer angle control provides a means to engineer band structure, transport and optical properties for developing tailored optoelectronic devices [8].

Beyond their remarkable material properties, 2D semiconductors have found widespread use in the field of light-matter coupling, particularly within both conventional and nanoscale optical cavities [19–21]. The large oscillator strength and tunable band structure make them ideal candidates for robust exciton-photon interactions, with implications for cavity quantum electrodynamics and polaritonic devices. This exceptional coupling behaviour has reignited interest in exciton-polariton physics, where the hybridization of excitons and photons leads to the formation of polaritons, quasi-particles with mixed light-matter properties [22]. In high-quality optical cavities, such as photonic crystals and microcavities, 2D materials can enter the strong coupling regime, where the coherent coupling of excitons and photons enables phenomena like Rabi oscillations and polariton condensation. On the other hand, nanoscale cavities, like plasmonic and dielectric nanoantennas, offer the ability to control light at sub-wavelength scales, facilitating the exploration of nonlinear and quantum optical effects. The ease of integrating 2D semiconductors into hybrid photonic structures expands the versatility of exciton-photon coupling in engineered platforms, as a result, 2D semiconductor-based photonic systems are advancing technologies such as low-threshold lasers, optical switches, and quantum light sources, positioning them at the forefront of nanooptics and optoelectronics innovation.

In this perspective, we provide a focused exploration of the cutting-edge research directions that leverage the unique properties of 2D materials in ultrafast and hybrid optoelectronic applications. In Section 2, we discuss moiré physics, highlighting recent advances in understanding the ultrafast excitonic and electronic dynamics that emerge in twisted 2D heterostructures. Section 3 focus on ultrafast magneto-optical studies, specifically examining spin and orbital phenomena in 2D materials. Finally, Section 4 addresses the dynamics at TMD/metal interfaces and provides a perspective on integrating 2D materials into hybrid photonic structures. This section emphasizes the opportunities that arise from strong exciton-plasmon interactions at these interfaces, which can enhance light-matter coupling and enable tunable optoelectronic responses. The investigation of ultrafast dynamics of light-matter interaction in hybrid 2D structures opens pathways for the design of novel quantum materials with customized optoelectronic properties. These insights aim to establish a framework for the continued development of 2D materials as versatile building blocks in ultrafast photonics, quantum technologies, and innovative optoelectronic devices.

## 2. Moiré systems

### 2.1 Optical properties and ultrafast dynamics of moiré semiconducting heterostructures

The discovery of unconventional superconductivity in twisted bilayer graphene superlattices revamped a great interest in 2D materials [23]. When the twist angle between the layers reaches about 1.1 degrees (the so-called 'magic angle'), the electronic band structure of twisted bilayer graphene exhibits the formation of flat bands near the Fermi energy, with correlated insulating behavior at half-filling, revealing a drastic change of the physical properties. The change of the twist angle between layers is associated with a modulation of the electronic wave function localizations due to the overarching electrostatic moiré potential, thus allowing to tune the physical properties, and opening the path to explore strongly correlated phenomena such as high-critical-temperature superconductors or quantum spin liquids.

The discovery of twisted graphene has shown the way to rethink on the properties of other 2D materials such as TMDs. In TMD heterobilayers (Fig. 1a), flat bands comparable to that of twisted bilayer graphene near "magic" angles have been theoretically predicted using density-functional theory to occur for 56.5° and 3.5° twist angles, with bandwidths of 5 and 23 meV, respectively [24]. The experiments on scanning tunneling spectroscopy (STS) on exfoliated TMD have confirmed the existence on moiré patterns in TMD heterostructures [25,26], with evidence of a strong 3D buckling reconstruction and indicating a dominant role of the strain redistribution on the electronic structure (Fig 1b-g). The STS imaging also demonstrate the existence of narrow and localized flat bands at the K-point of the band structure. The moiré band structure exhibits a complex structure, with series of flat bands and distinct degrees of spatial localization [25,26].

Regarding the optical properties of moiré superlattices in TMD, the exciton energy in moiré heterolayers depends on the twist angle due to interlayer hybridization, being the change remarkably strong for small twist angles [17,27]. Moreover, the valley physics of TMDs is also influenced. moiré heterolayers encode three quantum degrees of freedom like spin, valley index and layer index. Together with the modulation of the electronic and excitonic properties induced by the moiré patterns, we can confine spatially interlayer excitons with a locked spin-layer number [28]. In addition, playing with the carrier density and applied electric field we can control the nature of magnetic interactions [29], observe correlated insulating states at fractional fillings of the moiré minibands [30], and produce controlled metal-insulator transitions in twisted $WSe_2$ bilayers [31].

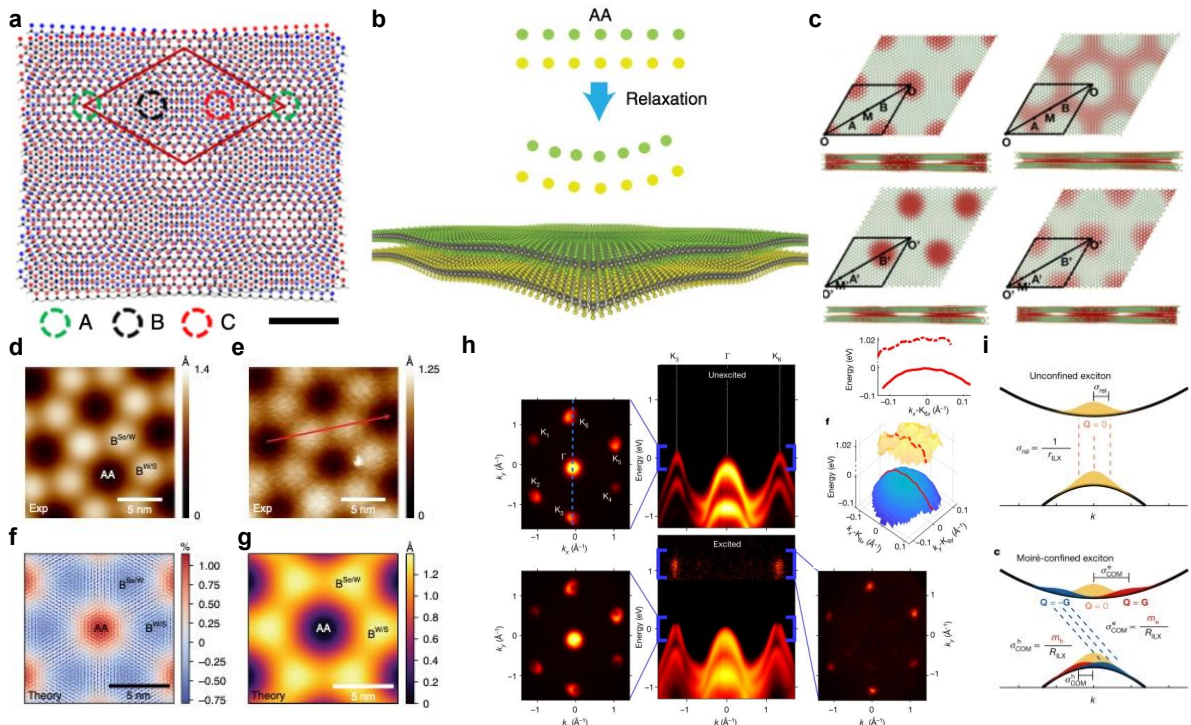

**Figure 1.** (a) Cartoon of the moiré superlattice formed in a $WSe_2/MoSe_2$ heterostructure with a twist angle of 5 degrees (scale bar 2 nm). Red diamonds represent a moiré supercell. Adapted from Ref. *[28]*. (b) Schematic of the buckling process in a TMD heterobilayer. Adapted from Ref. *[26]* . (c) Localization of electronic wave functions of the top of the valence band on moiré bilayer $MoS_2$. Adapted from Ref. *[24]* with permission of the American Physical Society. (d-e) Scanning Tunneling Microscopy images (STM) of $WSe_2/WS_2$ with a moiré period of 8 nm compared with a STM image of graphene-covered area. Adapted from Ref. *[26]*. (f-g) Theoretical in-plane strain distribution (f) for $WSe_2$ layer from DFT simulation and (g) theoretical height profile of the W atoms in the top of $WSe_2$ layer from DFT simulation. Adapted from Ref. *[26]*. (h) Energy–momentum cuts along the K–Γ direction without (top) and 25 ps after optical excitation (bottom). Lef: Momentum-space images without photoexcitation around VBM, as well as 25 ps after photoexcitation around VBM and around the ILX energy. Adapted from Ref. *[32]*. (i) Schematic of the momentum distributions of the electron and the hole constituting an unconfined exciton with zero center of mass momentum, showing identical extents. The black curves indicate the CB and the VB with their different curvatures. Adapted from Ref. *[32]*.

The rich physics arising in moiré heterolayers related to excitonic properties, mainly of TMD materials, open new venues to investigate ultrafast dynamics, especially with time-resolved angle resolved photoemission spectroscopy (TR-ARPES) [33]. For instance, the free carrier dynamics of single-layer $MoS_2$ was characterized for pump-probe experiments based on the excitation of free carrier into the conduction bands [34]. Only recently we learnt that excitonic features can be detected by TR-ARPES, probing the exciton formation and subsequent relaxation across the Brillouin zone [35]. The effects of phonon-induced decoherence and relaxation can be also investigated, and the excitation regime, either resonance or nonresonant with the excitonic states can be accessed [36]. Moreover, TR-ARPES signals including excitonic effects can now be simulated within fully ab initio methods [37].

Accompanying theoretical predictions, latest TR-ARPES data report a plethora of time-dependent phenomena with the exciton physics at the center of the optical properties. For instance, it is possible to observe the quasi-particle bandgap renormalization due to excitonic effects in monolayer $MoS_2$ [38]. Regarding TR-ARPES in moiré TMD, the dynamics of interlayer excitons is more elusive due to the large area and high-quality sample required. Experiments performed on $WSe_2/MoS_2$ heterostructure demonstrate the existence of interlayer excitons by capturing time-resolved and momentum-resolved distribution of electron and hole and even estimating with reliability the interlayer exciton diameter of 5.2 nm, comparable with the moiré unit cell of 6.1 nm [32]. In this work, the authors excite the moiré crystal with an optical pump of 1.67 eV in resonance with the A-exciton in $WSe_2$. Due to the type-II band alignment, the fast charge transfer results in the formation of the interlayer exciton, with the electron localized at $MoS_2$ and hole at $WSe_2$ (Fig. 1h-i). TR-ARPES provides electron and hole momentum distributions in an excitonic state, reporting an exciton behavior that resembles the one of exciton on impurity-based quantum dots and probing the exciton localization existing on the moiré materials. Recent femtosecond TR-ARPES measurements further investigated the formation of moiré excitons in $WSe_2/MoS_2$ with a twist angle of approximately 10° [39]. The study elucidates how interlayer excitons are formed, observing an intermediate femtosecond exciton-phonon scattering process into the dark Σ valley. Finally, the exciton Bohr radius is found to be approximately 1.6 nm, again comparable with the expected moiré periodicity of 1.8 nm.

## 2.2 Moiré in cavities

In semiconductor moiré superlattices a variety of excitonic species have been isolated, from the individual inter- and intra-layer exciton states arising from moiré induced confinement, to interlayer hybridization via excitons wavefunction-mixing (Fig. 2a) [6]. Such excitonic species have garnered significant attention in recent years for optoelectronic applications, owing to tunable optical properties and complex multi-body physics [17]. However, current optical measurements of these superlattices are constrained by the diffraction limit, which restricts the spatial resolution of far field experiments to the wavelength of the light probe. As a result, these measurements are only able to provide area-averaged information across more than 10,000 moiré cells, leading to reproducibility issues in moiré optical measurements and hinders the observation of intrinsic phenomena. Photonic structures, capable of confining light to ultra-small volumes, can overcome this limitation and enable optical investigations at scales commensurate with moiré periodicities [40].

Simultaneously, moiré-modulated interlayer coupling in bilayer materials presents an intriguing platform for exploring correlated Hubbard-model physics, which investigates strongly correlated phases of interacting bosonic particles confined in lattice potentials, like moiré excitons. Both transport and optical experiments have revealed distinct signatures of collective phases in these systems. Incorporating photonic elements into these systems holds the potential to manipulate these states, facilitating the optical control of quantum phases. For instance, optical cavities can lead to strong light-matter coupling regime and dressed-states, known as polaritons, where quantum vacuum (Rabi) fluctuations can be observed. With this approach, one could be driving the moiré system out of equilibrium to access hidden quantum phases that are not thermally accessible, even without an external driving force. This approach holds significant promise for realizing novel and high-temperature correlated non-equilibrium states [40] [].

While the focus of the research field has been mainly driven towards the realization of high quality devices, initial studies of moiré bilayer systems in optical cavities have been primarily focused on coupling the interlayer excitons to optical cavity modes [41], with demonstration of lasing regime in various configurations, from photonic crystals to nanobeam cavities [42,43]. In a recent work, strongly coupled interlayer excitons-polaritons with moiré induced nonlinearities have been reported in open optical microcavities based on distributed Bragg reflectors (DBRs) (Fig. 2b) [44]. Here, the strong coupling between moiré-confined excitons and cavity photons is revealed in the anticrossing between modes as observed in optical reflectance (Fig. 2c). By decreasing the photogenerated exciton density, it is found that heterobilayers exhibit large nonlinearities due to the exciton blockade in the moiré lattice cells (Fig. 2d-e), consistent with the theoretical picture of single occupancy model of the superlattice cells.

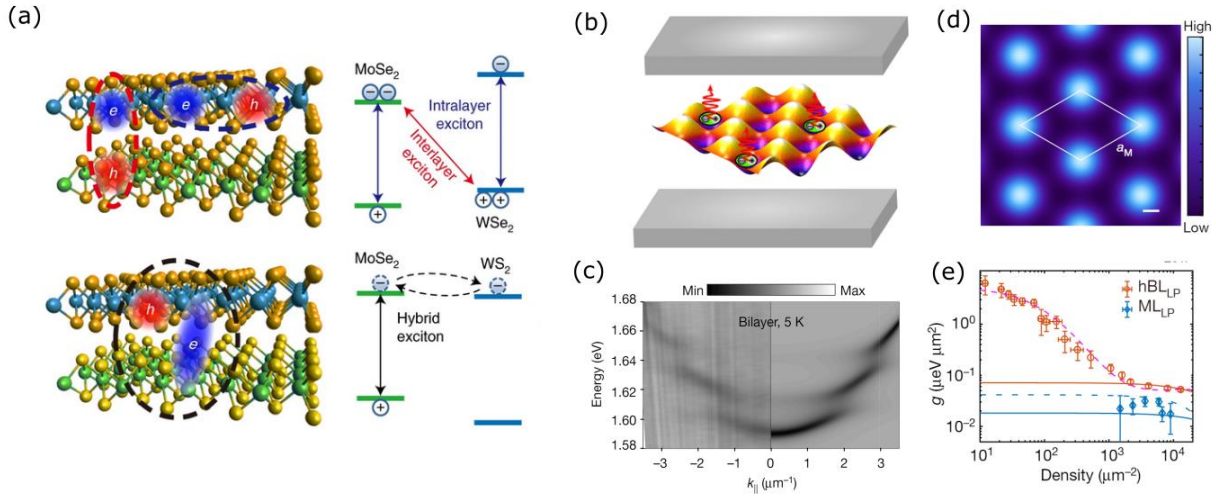

**Figure 2.** (a) Illustration of the interlayer (top) and hybridized (bottom) exciton species in TMDs heterobilayers. On the right, the respective band structure energy diagram. Adapted from Ref. [6]. (b) Illustration of the cavity coupling of moiré excitons. (c) Optical reflectance of a cavity coupled moiré TMD bilayer, exhibiting the anticrossing signature of the strong coupling regime. (d) Real-space distribution of the interlayer exciton component. The white line marks the moiré cell. Scalebar 1 nm. (e) Nonlinear coefficient (g) for a monolayer (blue) and heterobilayer (red) as a function of the photogenerated exciton density. The high nonlinearities at low densities are ascribed to the exciton blockade induced by the moiré potential. Panel b-e adapted from Ref. [44].

However, several challenges must be addressed to fully realize the potential of the photonic/2D moiré hybridized system. Theoretically, the coupling of a moiré system with photonic optical modes renders the original models used for understanding moiré physics inapplicable [45]. There is currently no clear strategy for solving these complex quantum Hamiltonians in 2D heterostructures using existing computational approaches. Thus, developing theoretical models for hybrid systems comprising photonic structures and moiré superlattices is essential for guiding research and enhancing the understanding of experimental results. Experimentally, photonic/2D moiré hybridized devices are highly sensitive to strain and defects that may be introduced during the integration of the 2D superlattices with photonic nanostructures. Additionally, the presence of such structures may induce fluctuations in the surrounding dielectric constant of the 2D moiré superlattice, potentially affecting the stability of the intrinsic quantum phases. Advances in material synthesis and nanofabrication techniques will be critical for developing high-quality photonic/2D moiré hybridized systems.

## 3. Ultrafast Magneto-Optics for Spin, Valley and Orbital Dynamics in Two-Dimensional Semiconductors and Heterostructures

Magneto-optical techniques are versatile, noninvasive and nondestructive tools to study various types of light-matter interaction at the nanoscale [46], in particular for spin and angular momentum dependent phenomena. For 2D materials, techniques such as the magneto-optic Kerr effect (MOKE) or magnetic circular dichroism give access to the spin, valley and magnetization degrees-of-freedom. This relies on the fact that opposite valleys (K and K') in semiconducting TMDs have different optical selection rules for circularly-polarized light (Fig. 3a). Therefore, a valley polarization can be achieved via light excitation with a certain helicity [47,48]. This change in valley occupation leads to strong magneto-optical signals, allowing one to detect the dynamics of valley and spin polarization through pump-probe time-resolved techniques. Due to the strong spin-orbit coupling in these materials, a valley accumulation is followed by a spin accumulation.

Time-resolved MOKE (TR-MOKE, Fig. 3a) and polarized-light time-resolved differential reflectivity (TRDR) have been widely used to study and understand the scattering mechanisms involved in the depolarization of the carriers – namely inter- and intra- valley scattering. The lifetimes extracted from these measurements reported in literature show a very large range of values, from a few ps to several ns [49–55]. Recent systematic experiments with different pump and probe energies, combined with electrostatic gating of the devices (Fig. 3b), point towards a major role for strain, dark states, and resident charge carriers on the longer lifetimes found in literature [56,57].

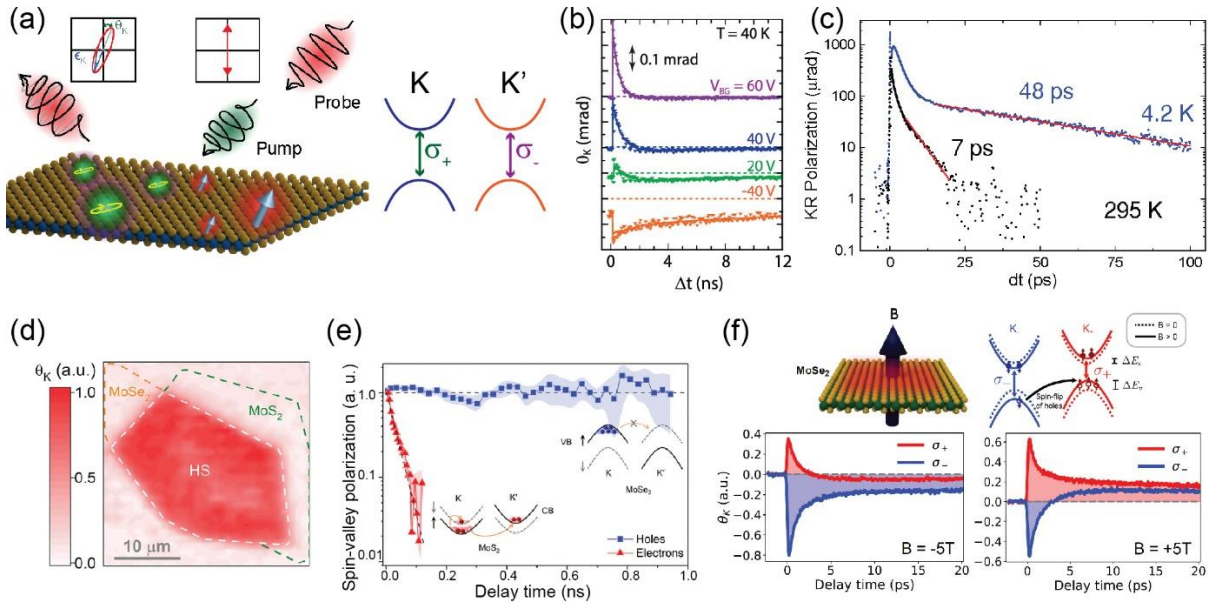

**Figure 3.** (a) TR-MOKE in a TMD monolayer. A circularly-polarized laser pulse (pump) generates a spin/valley accumulation. The valley excitation (K or K') is selected by different light helicities (right). The spin/valley accumulation is then probed by the polarization rotation or ellipticity of the reflection of a linearly-polarized laser pulse (probe). (b) TR-MOKE measurements in a $WSe_2$ monolayer showing gate-control over the spin lifetimes. Adapted from Ref. *[56]*. (c) Spin/orbital accumulation in bulk $WSe_2$ at 4.2 and 295 K. Adapted from Ref. *[58]*. (d) Energy-resolved spin/valley accumulation in a $MoSe_2/MoS_2$ heterostructure, with the pump exciting the $MoSe_2$ and the probe in resonance to the $MoS_2$ layer, with the pump-probe delay time at 2 ps. (e) Relaxation of electron and holes in $MoS_2$ and $MoSe_2$, respectively. Both (d) and (e) are adapted from Ref. *[59]*. (f) Control over the spin/valley lifetime using magnetic fields in $MoSe_2$ monolayers. The diagram of the valley-Zeeman energy shift is shown on top, and the TR-MOKE traces for positive and negative magnetic fields with the pump at different circular polarizations is shown on the bottom. Adapted from Ref. *[60]*.

Even though time-resolved magneto-optical techniques have been widely applied to monolayer TMDs, it is less true for devices consisting of bi- and few-layers, and heterostructures composed of different materials. Inversion symmetry – present in TMDs even layers and bulk – takes a state in one valley onto the other, however these states are also localized at opposite layers. As can be shown through a simple symmetry analysis, the states connected by inversion symmetry possess the same orbital character and therefore have the same optical selection rules [61]. Due to spin-orbit coupling, the states are also spin polarized in the same direction. Therefore, in the case of few-layer TMDs, circularly-polarized light leads to excitations with the same spin and the same orbital angular momentum but taking place at different valleys. This can then lead to a spin-orbital polarization under circularly-polarized light excitation. This is a generalized term, also applicable to monolayers. This optically-induced spin-orbital polarization has been experimentally and theoretically investigated in both bilayers and bulk TMDs (Fig. 3c) [58,61,62]. However, a predicted intervalley coherence and the electric control over its lifetimes through electrostatic gating remains to be demonstrated [63] and ultrafast magneto-optical techniques provide the ideal tool to do so. Finally, magneto-optics has been shown to be a powerful technique to explore the current-induced orbital polarization for the emerging field of "orbitronics" [64]. This was already explored in TMDs, where an electric current on a strained monolayer $MoS_2$ can induce a valley (orbital) polarization which can be measured through MOKE [65]. The relaxation mechanisms for this current-induced orbital polarization, however, remains unexplored and could lead to a deeper understanding on the scattering mechanisms responsible for valley depolarization at lower energy scales.

Heterostructures composed of different TMDs, such as $WSe_2/MoSe_2$, can host interlayer excitons [66], with long lifetimes, in the order of ns, allowing for an effective electric control over their resonance energies and dynamics using out-of-plane electric fields [67]. However, even though the interlayer exciton signatures in photoluminescence are very strong, their signatures in magneto-optical spectra are rather weak. This stems from their low oscillator strength for light absorption due to its spatially-indirect transition. To tackle this problem, one could either boost the light-matter interaction for interlayer excitons through the use of photonic structures [68], or explore thin-film interference effects to enhance the magneto-optical efficiencies [69] for these systems. Energy-resolved TR-MOKE experiments already demonstrated that spin/valley polarized charge carriers can be effectively transferred from one layer to the other (Fig. 3d) [59]. Additionally, electrons and holes have shown to possess dramatically different lifetimes in such heterostructures – namely $MoS_2/MoSe_2$ – with hole spin lifetimes extending well beyond 1 ns (Fig. 3e). The long lifetimes in combination with the possibility of electric tunability of light-matter interaction make TMD heterostructures highly promising for the on-chip electric manipulation of spin, orbital or valley information for applications in integrated (magneto-)photonic devices.

Proximity effects in heterostructures containing 2D magnetic materials, such as $CrBr_3$, provide a promising route to manipulate the spin/valley lifetime in TMDs. It has been shown that the photoluminescence of such structures shows a high degree-of-circular polarization which depends on the magnetization direction [70,71]. The origins of the photoluminescence polarization have been found to be from a spin-dependent charge transfer between the two materials. Similarly to what has been done for isolated TMD monolayers, time-resolved magneto-optical spectroscopy studies should help to understand if the proximity to $CrBr_3$ also leads to a polarization of (resident) carriers in the TMD and if it can be used to manipulate the spin lifetime in a nonvolatile fashion. Such manipulation of the spin lifetime through the valley-Zeeman effect has been demonstrated in $MoSe_2$ under high magnetic fields (Fig. 3f) [60], showing the potential of magnetic heterostructures for manipulating the spin/valley lifetimes. Additionally, these studies should help in the search for a combination of materials which would reduce the charge transfer – therefore improving the quantum efficiency for light emission – while also allowing for a strong magnetic proximity effect. Such a breakthrough would open an avenue for the stabilization of spin information for nonvolatile memory devices and information transfer.

## 4. Hybrid systems: interfaces and photonic platforms based on the integration of layered materials

### 4.1 Ultrafast dynamics at metal-layered semiconductor interfaces

The physical properties of vdW materials, such as TMDs, have being heavily investigated in the past decade due to their electronic and optical properties, promising for next-generation opto-electronic devices. In this context, unveiling the charge and spin dynamics, as well as energy transfer at the interfaces between either different TMDs or metals and TMDs, is crucial and yet quite unexplored [72]. In particular, understanding the ultrafast dynamics on the femtosecond timescale might offer new vibrant directions for the integration of photonics and electronics, especially in active and photo-detection devices [73], and more in general, in upcoming all-optical opto-electronic technologies.

As TMD-based devices typically utilize metals as contacts, it is crucial to study the properties of the TMD/metal interface, including the characteristics of the Schottky barriers [74] formed at the semiconductor-metal junction, also in view of future applications of the ultrafast (sub-ps) opto-electronic properties of TMDs. In this section, we focus on heterojunctions of metals and TMDs, since they have been proved to allow various possibilities for the manipulation and exploitation of light-matter interactions in a plenty of different research fields, such as plasmonics [75–81] and hot electron injection [82–92], transistors [93] and photovoltaics [94]. Furthermore, TMDs form atomically clean and sharp interfaces with other materials [14], which makes them ideal candidates for optoelectronic applications where high-quality interfaces between metals and semiconductors are essential, as well as they potentially offer a superior alternative to other semiconductors, as TMD/metal interfaces show weak Fermi-level pinning [95].

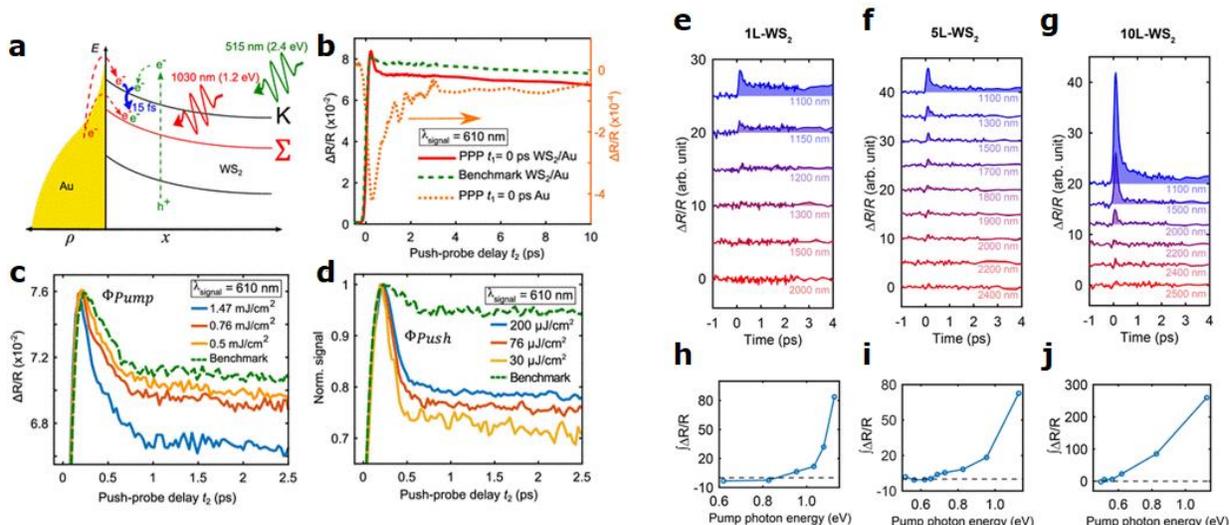

**Figure 4.** (a) Thermal Fermi-Dirac distribution (ρ) in gold and band alignment in $WS_2$ for the $WS_2$/Au heterojunction under illumination by pump pulse (red), followed by a modulated push pulse (green). The pump-induced thermionic injection of excess electrons (e–) from the gold and the direct excitation of free electrons (e–) and holes (h+) by the push pulse in WS2 is indicated by dashed arrows. The blue arrow indicates intervalley scattering, which migrates excited and injected electrons from the K to the Σ valley on a time scale of 15 fs. (b) Pump-push-probe measurement with pump-push delay of 0 ps on $WS_2$/Au (red line), bare gold (orange dotted line) and reference pump-probe measurement on $WS_2$/Au (green dashed line) at $\lambda_{probe}$ = 610 nm (2.03 eV). (c,d) Pump-push-probe curves at fixed pump–push delay of 0.1 ps at $\lambda_{probe}$ = 610 nm (2.03 eV) on WS2/Au for different pump ($\Phi_{pump}$) (c) and push ($\Phi_{push}$) (d) fluences. Adapted from Ref. [96]. (e-f) Pump-wavelength dependent ultrafast dynamics of (e) 1L-$WS_2$/Au, (f) 5L-$WS_2$/Au, and (g) 10L-$WS_2$/Au, extracted at the corresponding A exciton wavelengths. The ΔR/R kinetics are integrated over the first 4 ps for (h) 1L-WS2/Au, (i) 5L-$WS_2$/Au, and (l) 10L-$WS_2$/Au, with pump wavelength converted to pump photon energy in eV. Adapted from Ref. [97].

In this framework, a key aspect which has been investigated recently is the interplay between carrier injection and exciton formation dynamics at a van der Waals semiconductor/metal interface [98,96,99]. While a full theoretical description of the experiments performed recently by several groups has not yet been developed, it has been shown theoretically that an excess of free electrons in the conduction band of TMDs compared to the density of free holes affects the probability to form excitons and trions [100]. Also, experiments showing that an excess of electrons in the conduction band due to n-doping can modulate the excitonic absorption have been reported [101]. Furthermore, recent studies reveal that at $WS_2$/semimetal heterojunctions, hot carriers injected from the semimetal into a TMD (Fig. 4a) are able to affect the exciton formation dynamics by comparing the transient signal of pump-probe experiments for pumping above and below the optical band gap of the TMD [99,102,103].

To better understand this mechanism experimentally, it has been recently shown that a three-pulse pump-push-probe experiment allows to extract the effect of mutual interaction between injected and excited charge carriers on the transient signal in the absorption line of the exciton (see Fig. 4b-d). This approach enables to disentangle the effect of hot-electron injection from the metallic substrate from the direct excitations in the semiconductor, and this has helped to show how the ratio between thermionically injected and directly excited charge carriers affects the exciton formation dynamics in TMD/metal heterojunctions [96]. Also recently, it has been shown that by simply pumping below band-gap in similar heterojunctions to excite hot carriers from the metal to the layered semiconductor material with varying thicknesses it is possible to directly measure the Schottky barrier height and track the dynamics and magnitude of charge transfer across the interface (Fig. 4e-j) [97].

These versatile optical spectroscopy approaches for probing TMD/metal interfaces can thus shed light on the intricate charge transfer characteristics within various heterostructures, facilitating the development of more efficient and scalable nano-electronic and optoelectronic technologies. The open challenge here is to identify a strong theoretical framework to model the experimental results and allow a more proper design of the devices, as well as open new perspectives towards more exotic phenomena such as interlayer hybridization and the understanding of the spin and orbital degrees of freedom in different type of magnetic/Van der Waals materials interfaces.

## 4.2 Hybrid photonic structures

In recent decades, there has been a significant proliferation of diverse photonic structures, facilitating unprecedented control over light-matter interactions at both wavelength and subwavelength scales [104–106]. A new interdisciplinary field that utilizes photonic structures to engineer 2D material properties has also attracted increasing interest for advancing optoelectronic and quantum devices, as well as fundamental physics [107,108]. To control light-matter interaction in 2D materials there are two main options for photonic geometries: hybrid photonic devices and all-2D material photonic devices.

For the case of hybrid devices, these integrate single or few layers 2D materials with non-2D material photonic structures, such as plasmonic and dielectric nanoantennas [109,110], photonic crystals [111], or optical metasurfaces [112,113]. Due to the weak forces binding the layers to the substrate, vdW materials can be readily transferred to different photonic structures, making this the most used approach. Meanwhile, thanks to the pristine interfaces and ultrathin thickness, energy transfer between active 2D layers and passive photonic components can occur with high efficiency [114,115]. The main effect of photonic structures is that of altering the local density of states seen by the integrated 2D materials. This leads to either a weak light-matter coupling regime, with modification of the spontaneous emission rate, the so-called Purcell effect, or strong coupling regime where excitons and photons form new hybridized states, also known as polaritons. Whether the coupling is strong or weak depends on the exciton-photon coupling strength, proportional to the electric field intensity in the cavity and the exciton transition dipole moment, which requires to be larger than the intrinsic losses of the system [116,117].

For 2D materials in the weak coupling regime, enhancement or inhibition of photon emission via the Purcell effect [109,118,119] (Fig. 5a) and population inversion for lasing [120] (Fig. 5b) have been reported. In the strong coupling regime (Fig. 5c), instead, photons strongly couples with excitons [44,121,122], phonons [123], or mobile electrons [124] in 2D materials to form various half-light–half-matter polaritonic quasi-particles [125]. Owing to the strong light-matter coupling strength of monolayers TMDs, these have become a new standard for semiconductor-cavity coupling [126] with widespread demonstrations of exciton-polariton in various systems [20].

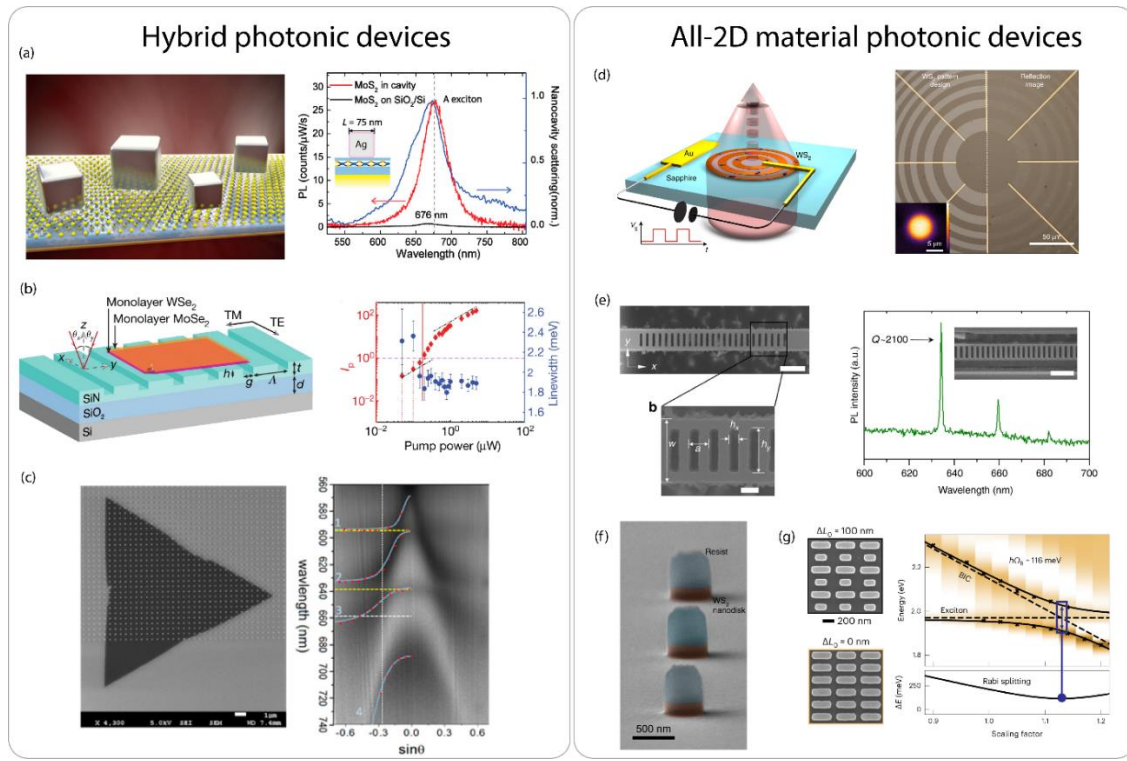

**Figure 5.** (a) 3D illustration of a plasmonic nanocavity, consisting of silver nanocubes of varying sizes on top of a gold film, and a $MoS_2$ monolayer (left). Enhanced PL emission from the $MoS_2$ monolayer observed in a nanocavity composed of a 75 nm nanocube (right). Adapted from Ref. [118]. (b) Schematic of the laser device, consisting of a heterobilayer on a grating cavity (left). The linewidth of cavity-mode emission shows a narrowing and a laser threshold intensity with varied pump powers (right). Adapted from Ref. [120]. (c) Strong exciton–plasmon coupling in silver nanodisk arrays integrated with a monolayer of $MoS_2$ (left). Angle-resolved differential reflectance spectra of 120 nm disk diameters of silver nanodisk arrays patterned on a monolayer of $MoS_2$ at 77K show anticrossing of dispersion curves near both A and B excitons (yellow lines), indicating strong exciton–plasmon coupling (right). Adapted from Ref. [127]. (d) Schematic (left) and optical microscope image (right) of an atomically thin zone plate lens. A clear focal spot is formed approximately 2 mm above the patterned surface ($\lambda$ = 620 nm) behind the lens, with a characteristic flat-top beam profile and a full-width at half-maximum of 6.7 μm (right inset). Adapted from Ref. [128]. (e) SEM image of one-dimensional (1D) hBN photonic crystal cavities (left). The PL spectrum of a 1D cavity fabricated by focused ion beam milling shows a high Q (~2100) mode in the visible spectral range (right). Adapted from Ref. [129]. (f) Side-view SEM images of fabricated $WS_2$ nanodisks. Adapted from Ref. [130]. (g) SEM imaging of symmetric and asymmetric $WS_2$ metasurfaces sustaining quasi bound states in the continuum (qBIC) resonances (left). Optical transmittance of a set of qBIC $WS_2$ metasurfaces where the anticrossing between the exciton and the metasurface/cavity qBIC resonance confirms the achievement of strong coupling regime in ambient conditions. Adapted from Ref. [131].

On the other hand, all-2D material photonic devices involve nanofabricating single or multilayered exfoliated materials themselves into predesigned optical nanostructures, able to modify the interaction of light employing the intrinsic optical properties of the layered 2D material [130,132,133]. In particular, the high binding energy of 2D semiconductor indicates higher stability up to ambient conditions, and a high refractive index, allowing for ultrathin photonic devices [128,134]. Optically thin metalenses have been realized in TMD monolayers [135], with demonstration of electrical tunability driven by the excitonic resonances (Fig.5d) [128].

Beyond the single layer form, which unable to confine light as in conventional optical cavities due to the atomical thickness, multilayered 2D materials have emerged as a new platform for integrated photonics [129–131]. The vast library of van der Waals materials offer an unprecedented opportunity to investigate and develop novel photonic platform. For instance, hexagonal boron nitride (hBN) owing to its transparency region over the visible range can be nanostructured into suspended membranes to achieve photonic crystal cavities with high Q factors (Fig.5e) [136]. Similarly, high Q factor metasurfaces in the visible have been demonstrated by leveraging the physics of quasi-bound states in the continuum (qBIC) [129]. In the case of TMDs, their bulk form has gathered wide attention for optical applications, beyond the intrinsic stable excitons, owing to high refractive indexes (n>4) and record high optical anisotropies [137–139].

The initial demonstration of all-TMD nanophotonic structures for achieving strong light-matter coupling was realized in $WS_2$ nanodisks (Fig.5f), where anapole resonances were shown to couple effectively with excitons [133,140]. This pioneering approach demonstrated the feasibility of exciton-resonance coupling with intrinsic optical resonances in TMD materials, albeit with limitations in Q factors and design flexibility. In fact, the geometry of $WS_2$ nanodisks, while effective for initiating strong coupling, lacks the tunability needed for more optimized and application-specific resonances. Recent advancements have addressed these challenges through the development of high-Q factor metasurfaces leveraging qBIC physics, which offer significantly improved light-matter interaction [141,142]. These ultrathin qBIC metasurfaces, engineered for enhanced resonance control by controlling the design of the asymmetric unit cell geometry, enable robust exciton-photon coupling in $WS_2$ thin layers (Fig.5g), with coupling strength above 55 meV at room temperature [143]. This evolution to high-Q factor metasurfaces highlights the progress towards creating more versatile and scalable platforms, paving the way for more practical and efficient nanophotonic devices that operate effectively at ambient conditions.

Finally, the potential for chiral light-matter coupling in 2D materials remains largely unexplored. Introducing chirality into cavity-coupled systems could enable control over valley dynamics, providing a novel mechanism to influence spin and valley polarization in TMDs. By leveraging chiral coupling, it may be possible to selectively separate chiral emissions from the sample, opening pathways for chirality-dependent or independent light propagation [97]. Such an approach could significantly advance the control of valleytronic processes, offering exciting new directions for modulating and harnessing chiral properties in 2D materials for applications in quantum information and photonic devices. By selectively coupling with valleys, enhanced chiral emission [131] and chiral lasing [144] have been reported. Certain photonic structures, such as waveguides [145,146], asymmetric groove array [147] and plasmonic nanowires [148], can also be used to direct emission from valley polarization excitons toward different directions, effectively serving as a link between photonic and valleytronic devices.

## 5. Outlook

In conclusion, 2D materials, particularly TMDs, offer a remarkable platform to explore and harness ultrafast excitonic, spin-valley, and moiré physics. Their unique quantum properties, strong light-matter interactions, and ease of integration in hybrid photonic structures enable the engineering of new optoelectronic phenomena and devices, from valleytronics-based memory devices and low-threshold polariton lasers to tunable quantum simulators based on moiré superlattices and van der Waals heterostructures. Furthermore, their ultrafast response dynamics and compatibility with diverse interfaces provide a compelling advantage for next-generation photonic and quantum devices, and a florid ground for fundamental research. The exploration of proximity effects and moiré potentials show a unique opportunity for the engineering and manipulation of spin, valley and orbital degrees of freedom, which can be studied through optical experiments. This provides an ideal test ground for testing theoretical predictions and the exploration of new physics in, for example, the new field of orbitronics. By further integrating photonic structures with 2D materials, photons can weakly or strongly couple with 2D material, driving it into a non-equilibrium state and enabling the development of novel optoelectronic and opto-quantum devices. Expanding on this approach to control collective phenomena in moiré systems with photonic structures could open new avenues for exploring many-body physics in solid-state systems. The exploration of these materials as foundational building blocks of layered architectures paves the way for breakthroughs in ultrafast photonics, optoelectronics, and quantum information science.

## Acknowledgements

NM acknowledges support from the Swedish Research Council (Grant No. 2021-05784), the Knut and Alice Wallenberg Foundation through the Wallenberg Academy Fellow Program (Grant No. 2023.0089) and the European Innovation Council (Grant No. 101046920). MHDG acknowledges the support by the European Research Council (ERC, 2D-OPTOSPIN, 101076932). AMS acknowledges support from the European Union's Horizon Europe research and innovation program under the Marie Sklodowska-Curie grant agreement 101118915 (TIMES), project I+D+i PID2023-146181OB-I00 UTOPIA, funded by MCIN/AEI/10.13039/501100011033, project PROMETEO/2021/082 (ENIGMA) and SEJIGENT/2021/034 (2D-MAGNONICS) funded by the Generalitat Valenciana. This study is also part of the Advanced Materials program (project SPINO2D), supported by MCIN with funding from European Union NextGenerationEU (PRTR-C17.I1) and by Generalitat Valenciana. JQ acknowledges the support of the Shanghai Pujiang Talent Plan (Grant No. 23PJ1413100) and the Rising-Star Program (Grant No. 24QA2709500).